\documentclass[preprint]{aastex}

\newcommand{\Ms}{\ensuremath{M_{\odot}}}
\newcommand{\eg}{{\it e.g.}}
\newcommand{\cf}{{\it c.f. }}
\newcommand{\ie}{{\it i.e.}}
\newcommand{\viz}{{\it viz.}}

\slugcomment{To appear in The Astrophysical Journal}

\shorttitle{Runaway massive stars from R136}
\shortauthors{Banerjee, Oh \& Kroupa}

\begin{document}


\title{Runaway massive stars from R136: VFTS 682 is very likely a ``slow runaway''}


\author{Sambaran Banerjee, Pavel Kroupa and Seungkyung Oh}
\affil{Argelander-Institut f\"ur Astronomie, Auf dem H\"ugel 71, D-53121, Bonn, Germany}

\email{sambaran@astro.uni-bonn.de, pavel@astro.uni-bonn.de, skoh@astro.uni-bonn.de}


\begin{abstract}
We conduct a theoretical study on the ejection of runaway massive stars from R136 --- the
central massive, star-burst cluster in the 30 Doradus complex of the Large Magellanic Cloud.
Specifically, we investigate the possibility of the very massive star (VMS) VFTS 682 being a
runaway member of R136. Recent observations of the above VMS,
by virtue of its isolated location and its moderate peculiar motion,
have raised the fundamental question whether isolated massive
star formation is indeed possible. We perform the first realistic N-body computations of fully
mass-segregated R136-type star clusters in which all the massive stars are in primordial binary systems.
These calculations confirm that the dynamical ejection of
a VMS from a R136-like cluster, with kinematic properties similar to those of VFTS 682, is common.
Hence the conjecture of isolated massive star formation is unnecessary
to account for this VMS. Our results are also quite consistent with the ejection of 30 Dor 016, another
suspected runaway VMS from R136. We further note that during the clusters' evolution, mergers of massive binaries
produce a few single stars per
cluster with masses significantly exceeding the canonical upper-limit of $150\Ms$. The observations
of such single super-canonical stars in R136, therefore, do \emph{not} imply an IMF with an upper limit greatly exceeding
the accepted canonical $150\Ms$ limit, as has been suggested recently, and they are consistent with the canonical upper limit. 
\end{abstract}





\section{Introduction}\label{intro}

Fast-moving massive field stars in our Galaxy and in external galaxies, that apparently are not
associated with any nearby stellar assemblies, have been a focal topic for the past two
decades. Such runaway OB stars are widely thought to be members of star clusters that
are ejected following dynamical encounters in their parent clusters. If the proper
motions of these massive stars can be reliably estimated, the vast majority of them can then be traced back to
their parent clusters. Using this kinematic method, \citet{silros2008} have demonstrated that
most of the Galactic OB runaway stars indeed originated from star clusters.
Another way to determine the parents of runaway massive stars
is to image their bow shocks, if present, which has been possible with modern infrared
telescopes \citep{gv2010,gv2011}. A detectable bow shock is generated if a star moves supersonically
(typically $\gtrsim 10$ km s$^{-1}$) through a dense enough medium.
The geometry of a bow shock allows one to determine the
direction of motion of the shock-generating star and hence its possible parent cluster \citep{gv2010}.
Bow shocks are particularly useful when the proper motions of the runaway OB stars are
unavailable or are poorly known which is often the case for distant stars. However, even if every OB field
star has been ejected from a young cluster, 1-4\% of the OB stars cannot be traced back to their
parent clusters because of the two-step ejection process \citep{pfl2010}: a massive binary is first ejected from its
cluster and when the primary explodes as a supernova, the secondary OB star continues on a
diverted trajectory.

High-velocity single and binary stars are launched from a star cluster due to
super-elastic single-star---binary and binary-binary dynamical encounters \citep{hg75} that occur
predominantly in the cluster's dense core. Therefore, the spectrum of the ejected bodies
is governed by the properties of the primordial binaries in the cluster \citep{ld88,cp92} and its
stellar initial mass function (IMF) and hence serves as a useful tracer of these fundamentally important cluster
properties. As the stars and the binaries are mostly ejected from the cluster's
core, where the most massive members remain segregated (mass stratification; see \citealt{spz}), the
runaway population contains a large fraction of OB single stars/binaries. The ejection of
the massive members also alters the stellar mass function (MF) of the bound cluster as it
evolves.

A particular runaway very massive star (VMS) that has recently drawn significant attention is
VLT-FLAMES Tarantula Survey (VFTS) 682.
This VMS, visible in the Tarantula Nebula (30 Doradus) of the Large Magellanic Cloud (LMC), has an inferred
present-day mass of $\approx 150\Ms$ and is located at a projected distance of $\approx 30$ pc
in the North-East from the central massive star cluster R136 of the 30 Doradus complex \citep{blh2011}.
From radial velocity measurements and its projected separation from R136, \citet{blh2011} inferred  
that the true velocity of VFTS 682 should be $\approx 40$ km s$^{-1}$ if it is
indeed a runaway from R136. On the other hand, the apparent isolation of VFTS 682
from any star cluster, its moderate peculiar motion and the non-detection of a bow shock
(but see \citealt{gv2010}) might also indicate that this VMS has formed isolated and is unrelated to R136,
as \citet{blh2011} argue. The origin of VFTS 682 is therefore currently unclear, \ie, whether it is
a ``slow runaway'' from R136 or has formed alone, outside any clusters. This, in turn, freshens up the open question
whether massive stars form only in dense environments \citep{bonn2004} or whether isolated massive star formation
is possible through pure circumstellar disk-accretion \citep{kui2010}.
Notably, the stellar content of R136 itself presents a challenge
to the current understanding of star-formation phenomenology as it contains several
single-star members with inferred initial masses upto $\approx 320\Ms$, \ie, well above the widely accepted
$150\Ms$ canonical upper-limit of the stellar IMF \citep{crw2010}. As coined by \citet{pk2011}, the stellar population
of R136 is therefore ``super-saturated'' by containing super-canonical stars.

Because of the important implications in massive star formation theory, it is essential to determine
the likeliness of a VFTS 682-like runaway from R136. To that end, we compute the evolution of
model star clusters with properties resembling R136 using the direct N-body integration method and focus on
the kinematics of the massive stars ejected from them. We find that the ejection of
VMSs, within a few Myr, with kinematic properties similar to VFTS 682 is quite plausible: the latter star is then
quite possibly a runaway from R136 and the substantially controversial assumption of isolated massive star
formation is unnecessary to explain its existence. These computations also demonstrate that super-canonical
stars appear naturally from merged binaries.

In Sec.~\ref{comp}, we describe our model clusters (Sec.~\ref{initcond}) and the method of our
computations (Sec.~\ref{nbint}). In Sec.~\ref{res} we present our results focussing on
VMS runaway members that resemble VFTS 682 (Sec.~\ref{vfts682}) and also on 30 Dor 016-like
runaways (Sec.~\ref{dor16}). We also discuss how the ejected fraction of stars depends on the
stellar mass (Sec.~\ref{ejctfrac}). We conclude this paper in Sec.~\ref{discuss} with a discussion
and highlighting possible future improvements.

\section{Computations}\label{comp}

\subsection{Initial conditions}\label{initcond}

As our primary objective is to study the runaway massive stars from a R136-like cluster, we begin
our N-body computations with star clusters modelled as Plummer spheres \citep{pk2008} having properties
conforming with the observed parameters of R136. The present-day parameters
of R136 still remain ambiguous. We take the total initial mass of each of our model
Plummer spheres to be $M_{cl}(0) \approx 10^5\Ms$ which is an upper limit of the mass
of R136 \citep{crw2010}. As for its size, a variety of half-mass radii/core radii has so far been assigned to this
cluster by different authors \citep{hunt95,pz2002,mg2003,and2009,pz2010}.
We take the initial half-mass radii of our
models to be $r_h(0) \approx 0.8$ pc. The core radii of our models then turn out to be
$r_c \lesssim 0.3$ pc throughout their evolutions, 0.3 pc being an observed upper limit \citep{mg2003}.
The core density is $\gtrsim 1.3\times 10^4$ stars pc$^{-3}$.

The initial masses $m_s$ of the clusters' member stars are drawn from the canonical IMF \citep{pk2001}
in the range $0.08\Ms < m_s < 150\Ms$. This IMF is a two-part power-law with indices $\alpha_{1} = 1.3$ for
$0.08\Ms < m_s < 0.5\Ms$ and $\alpha_{2} = 2.3$ for more massive stars. Throughout this paper,
we denote the (instantaneous) mass of a stellar member by $m_s$ in general and $M$ is exclusively reserved to denote
the mass of a runaway (see Sec.~\ref{res}) member. The metallicity of the
stars are taken to be low: $Z=0.5Z_\odot$ which is appropriate for R136.

Most stars form in binaries or higher-order multiple systems. Furthermore, close binary-binary
encounters play an important role in the dynamical ejection of stars from a cluster
\citep{ld88,ld90}. However, at the current state-of-the-art,
binaries are computationally the most time-consuming members in a direct N-body calculation of a star cluster
because of the particular numerical algorithms involved (see Sec.~\ref{nbint}).
Thus, in order to achieve a reasonable calculation-speed,
we initially arrange all the stars more massive than $5\Ms$ in binaries
and all the less massive members are kept single.
As our primary aim is to probe the \emph{massive} runaway stars, which can be efficiently
ejected only via close encounters with massive, hard binaries (see Sec.~\ref{vfts682}), only massive binaries are important
in this study.

To create the initial binary population, we first sort the $m_s > 5\Ms$ stars with 
decreasing mass and then pair them in order, so that the resulting binaries have their 
mass-ratios close to unity. It is already understood
that a mass-ratio distribution biased towards unity is required in order to get a significant ejection rate 
of massive stars \citep{cp92}. Furthermore, observations indicate that 
O-stars favor OB stars as their companions \citep{sev2010}.
The orbital periods, $P$, of the binaries with primary masses $ > 20\Ms$ are chosen 
from a uniform distribution between $0.5 < \log_{10} P < 4$ (\"Opik's law; $P$ in days)
since the O-type stars are preferentially found in short-period systems.
This range of $P$ is similar to that obtained by \citet{sev2010}.

For $5\Ms < m_s < 20\Ms$, equation (8) of \citet{pk95b} is chosen
as the period distribution, which is given by
\begin{eqnarray}
f_{P,\rm birth} = \eta\frac{\left(\log_{10}P - \log_{10}P_{\rm min}\right)}
{\delta + \left(\log_{10}P - \log_{10}P_{\rm min}\right)^2}, \nonumber
\end{eqnarray}
with $\eta=2.5$, $\delta=45$ and which covers a much wider range of period between
$1.0 < \log_{10} P < 8.43$ \citep{pk95a,pk95b,mrk2011}.
Here, $f_{P,\rm birth}$ is the ``birth period distribution'' of binaries as obtained
by \citet{pk95a} through ``inverse dynamical population synthesis''. It is the primordial
binary period distribution unmodified by the dynamical destructions of these binaries
(\ie, without a depletion function; \citealt{pk95a}) or by the mutual interactions between the binary
members in their pre-main-sequence stage (\ie, without an ``eigenevolution''; \citealt{pk95b}).
The modification of a primordial binary's orbital period, mass-ratio and eccentricity following its
eigenevolution is known only for low mass stars \citep{kgot2011,mrk2011}.
We choose the orbital eccentricities $e$ following a thermal distribution,
\ie, $f(e) = 2e$ \citep{spz,pk2008}. Fig.~\ref{fig:EvM} shows the resulting distribution of binary binding
energies as a function of the primary mass. The two distinct energy distributions across $m_s = 20\Ms$
are clearly visible; the binaries with $m_s > 20\Ms$ primaries are harder
than those with $m_s<20\Ms$ primaries.

The initial systems are completely mass-segregated \citep{spz} using
a method by \citet{bg2008} so that the heaviest
stars are initially concentrated within the clusters' cores. Such an initial
condition mimics primordial mass segregation that is inferred to be true for several Galactic globular
clusters \citep{bg2008,mrk2010,zooz2011,hasan2011} and young star clusters \citep{lit2003,chen2007}.
Notably, the initial setup employed
here (all massive stars in binaries, fully mass-segregated cluster) are the very
first ones ever computed under such extreme conditions.

\subsection{N-body integration, stellar evolution and
stellar hydrodynamics}\label{nbint}

The initial models, constructed as above, are dynamically evolved  using the state-of-the-art
direct N-body integrator ``NBODY6'' \citep{ar2003} which takes advantage of the remarkable hardware
accelerated computing capacity of a Graphical Processing Unit (GPU) while integrating the
centers of masses. Algorithmically, the
most important feature of NBODY6 is that it applies regularization techniques \citep{ar2003} to resolve
close encounters, making them highly accurate and therefore uniquely suitable for this work.
However, when there are a significant number of primordial binaries, their regualrized orbits
can currently be integrated only on the much slower host workstation processors, which bottlenecks
the GPU's hardware acceleration significantly. No external tidal field is applied during the
N-body integration as the gravitational field structure of LMC is currently unclear.  

In addition to integrating the point-mass motion, NBODY6 also employs analytical but
well-tested stellar evolution recipes \citep{hur2000} to evolve the individual stars. 
These prescriptions are based on model stellar evolution tracks computed by
\citet{pols98}. The wind mass loss of all massive stars are taken into account
using the empirical formula of \citet{nj90}, while they are on the main sequence.
The mass loss rates are, of course, appropriately modified on the giant branches and other evolved
phases for stars of all masses (see Sec.~7 of \citealt{hur2000}). 
The code incorporates the physics of stellar binary evolution as well \citep{hur2002}.
It also includes
detailed models for tidal interactions between stars and recipes for the outcomes of
mergers between different types of stars and stellar remnants (see Table 2 of \citealt{hur2002}
for a summary).

Since mergers among main-sequence (hereafter MS) stars have substantial consequences in our results (Sec.~\ref{res}), we
summarize here the treatment of a merger between two MS stars in NBODY6 (which follows 
\citealt{hur2002} schemes; see also \citealt{hur2005}).
When two MS stars collide (a collision between two single MS stars or between two MS components
in a binary due to eccentricity induced by close encounters and/or due to encounter hardening),
it is assumed that (a) the merged product is a MS star with the
stellar material completely mixed and (b) no mass is lost from the system during the merger
process. The no mass-loss assumption is based on the results of SPH simulations of MS-MS
mergers (\eg, \citealt{sil2001}) but such calculations yield only a limited amount of mixing. The rejuvenated
age of the merged MS star is determined depending on the amount of unburnt hydrogen
fuel gained by the hydrogen-burning core as a result of the mixing. In case of a
mass transfer across a MS-MS binary, distinction is made between the cases when the original accretor MS star
has a radiative or a convective core. For a convective core, the core grows
with the gain of mass and mixes with the unburnt hydrogen fuel so that the accreting MS
star appears younger. For the case of a radiative core, the fraction of the hydrogen burnt
in the hydrogen-burning core remains nearly unaffected by the gain of mass
so that the effective age of the MS star decreases. The effective age
is determined so as to keep the elapsed fraction of its MS lifetime unchanged.

The initial cluster mass of $M_{cl}(0) \approx 10^5\Ms$ in our models corresponds to $N(0) = 170667$
stars. To our knowledge, direct N-body computations with such a large number of stars,
where the clusters are fully mass-segregated and all the massive stars are in binaries, are being reported
for the first time. We evolve 4 initial models with the above $N(0)$, generated using different
random number seeds, until $\approx 3$ Myr. We take this age as an upper limit of the age of R136 \citep{crw2010,pz2010}.
We do all the computations on ``NVIDIA 480 GTX'' GPU platforms.

\section{Results}\label{res}

From the above computations, we trace the bodies that are ejected from the clusters during
their evolution (within $\approx 3$ Myr). We consider a single-star/binary/multiplet to be a runaway member from its host
cluster if it is found moving away from the cluster beyond $R>10$ pc distance from the cluster's center of density.
Although our primary focus is on the runaway VMSs, we consider the whole mass spectrum
of ejected stars as well. 

\begin{figure}
\centering
\vspace{-3.5cm}
\includegraphics[angle=0,height=15cm]{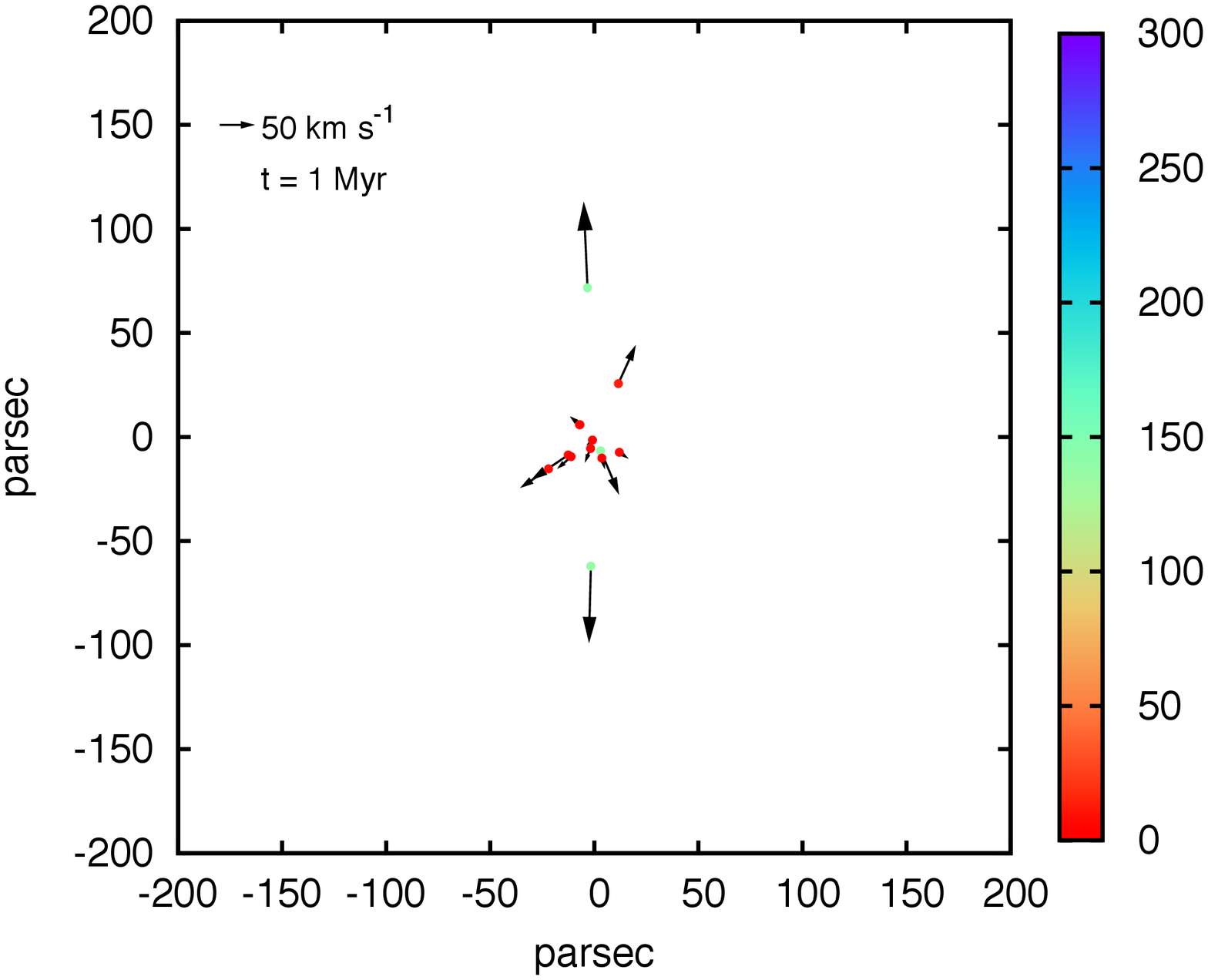}\vspace{-6.5cm}
\includegraphics[angle=0,height=15cm]{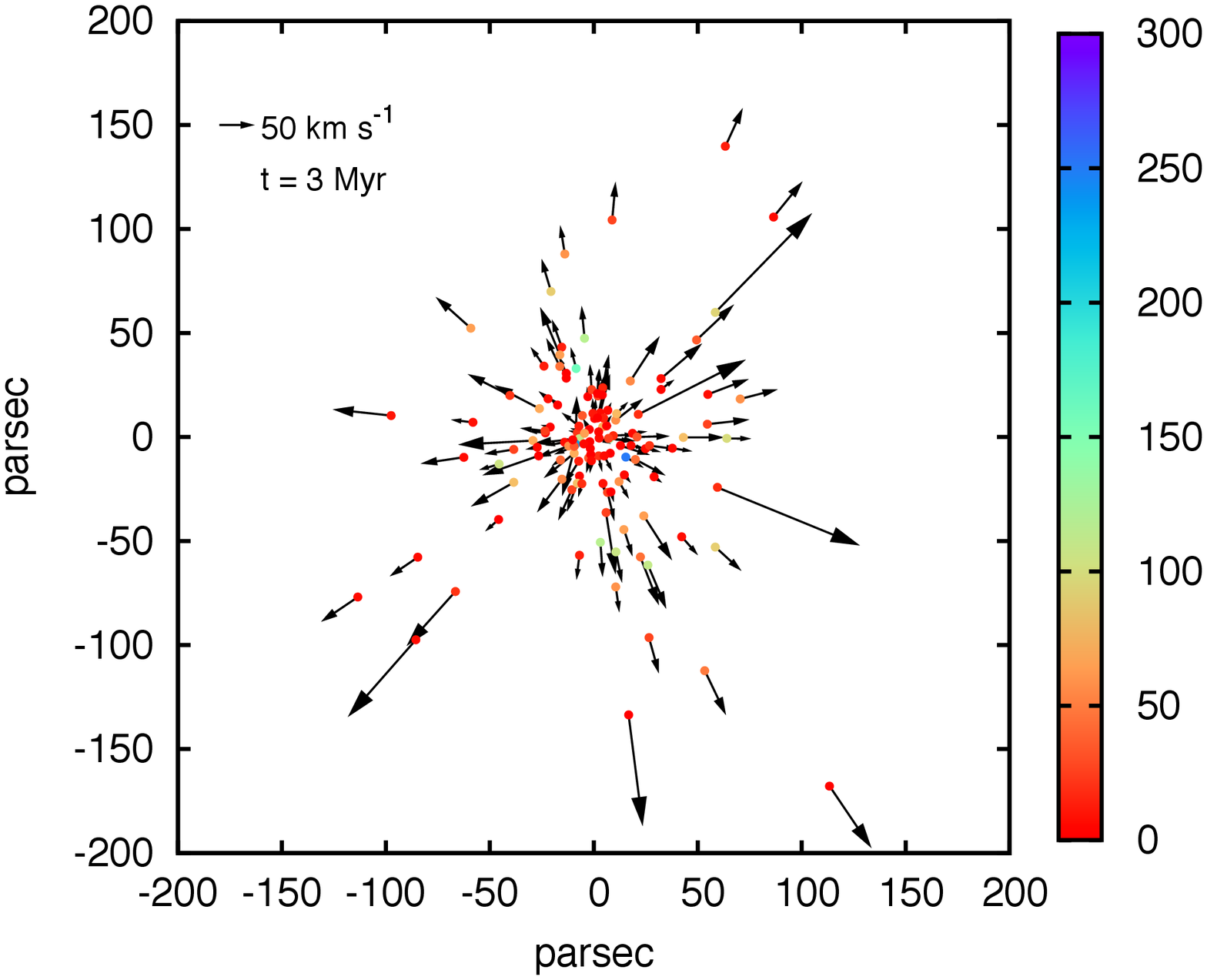}\vspace{-3.9cm}
\caption{Snapshots of runaway stars (a single filled circle) with $M>3\Ms$ at
$t=1$ Myr (top) and 3 Myr (bottom) evolutionary times as obtained
from our computations.
The snapshots from the 4 computations are superimposed in each panel. The stars
are colour-coded according to their masses at age $t$, the values in the colour code (colour-bar on the
right hand side) being in $\Ms$. The direction of the arrow that originates from each star gives
its direction of motion in the plane of the snapshot while its length is proportional to
the 3-dimensional velocity of the star, the scale being shown at the upper-left corner of each panel.
It can be seen that there are a significant number
of fast runaway VMSs by $t=3$ Myr with 3-dimensional velocities reaching upto $\approx 300$ km s$^{-1}$ 
and a few fast VMSs are already present at $t=1$ Myr (also, see Fig.~\ref{fig:mvplot}). See text for details.}
\label{fig:ejsnap}
\end{figure}

Fig.~\ref{fig:ejsnap} shows the projected snapshots of the runaways with
masses $M>3\Ms$, combined from the 4 computations, at $t=1$ Myr and 3 Myr evolutionary times. It can be
seen that by $t=3$ Myr there are a significant number of fast runaway VMSs with total (3-dimensional) velocities
upto $\approx 300$ km s$^{-1}$ and a few fast VMSs are already present at $t=1$ Myr. All these
runaways are on their MSs and hence are OB stars. We note that the vast majority
of the massive ejected members are single stars --- only 2 of the massive ejecta from our computations are found
in hard binaries. We shall discuss the multiplicity properties of the ejected stellar population in detail
in a future paper. 

It is worthwhile to note that although our adapted canonical IMF has a $150\Ms$ upper limit, our models yield single-star runaways
with masses upto $\approx 250\Ms$ within $t<3$ Myr, considerably exceeding the widely accepted $150\Ms$
limit. A few of such members are also found bound to each of our model clusters within the same
evolutionary period. These VMSs form when massive binary components merge as a result of encounter hardening \citep{hg75,bg2006}
of these binaries and/or eccentricities induced to them by close encounters. It is the adapted relatively small
and narrow orbital period distribution of the massive binaries ($m_s>20\Ms$), conforming with the observed properties of O-type 
stellar binaries (see Sec.~\ref{initcond}),
that makes such mergers probable, leading to the formation of single stars with $m_s > 150\Ms$.
On the other hand, it is these hard, massive binaries that can efficiently drive the VMS
runaways (see Sec.~\ref{vfts682}). In other words, the possibility of the presence of runaway VMSs
naturally leads to the formation of VMSs (bound or ejected),
in the course of the evolution of the cluster, with masses significantly exceeding the canonical upper limit,
even though the cluster's IMF is intrinsically canonical. Therefore, the observed super-canonical VMSs in R136,
with inferred individual 
initial masses upto $\approx 320\Ms$, do not necessarily imply an IMF with an upper-limit greatly exceeding the canonical limit as
argued by \citet{crw2010}: the $m_s > 150\Ms$ single-star members can as well be accounted for as
recent massive binary merger products.

Each of our computed models yields a few of such super-canonical single-star members. Because of their substantial
gravitational focussing \citep{spz}, they are involved in very frequent close encounters with other massive
binaries, thereby being vulnerable to being ejected from the cluster. Three of our 4 computed clusters
have produced super-canonical single-star runaways with $M > 150\Ms$ and the most massive merger product has been
ejected from 2 of the models.

\subsection{VFTS 682: a very likely runaway from R136}\label{vfts682}

To have an understanding of the spectrum of velocities with which the stars run away, we plot their
3-dimensional velocities $V$ as a function of their instantaneous masses $M$ at $t=1$ Myr and 3 Myr
as shown in Fig.~\ref{fig:mvplot}. The data points from all the computations have been compiled in this figure,
which are distinguished by the different symbols. We note two trends in this plot, \viz,
(a) there is a lower boundary of the scatter in $V$ that increases moderately with $M$
and (b) the upper boundary of this scatter is independent of $M$. 
These trends are vividly apparent from the plot at $t=3$ Myr (Fig.~\ref{fig:mvplot}, lower panel) where there is a significantly
larger number of runaways. No well-defined boundaries can, of course,
be drawn in this plot and the above mentioned boundaries are meant to indicate net trends.

\begin{figure}
\centering
\includegraphics[angle=0,height=9cm]{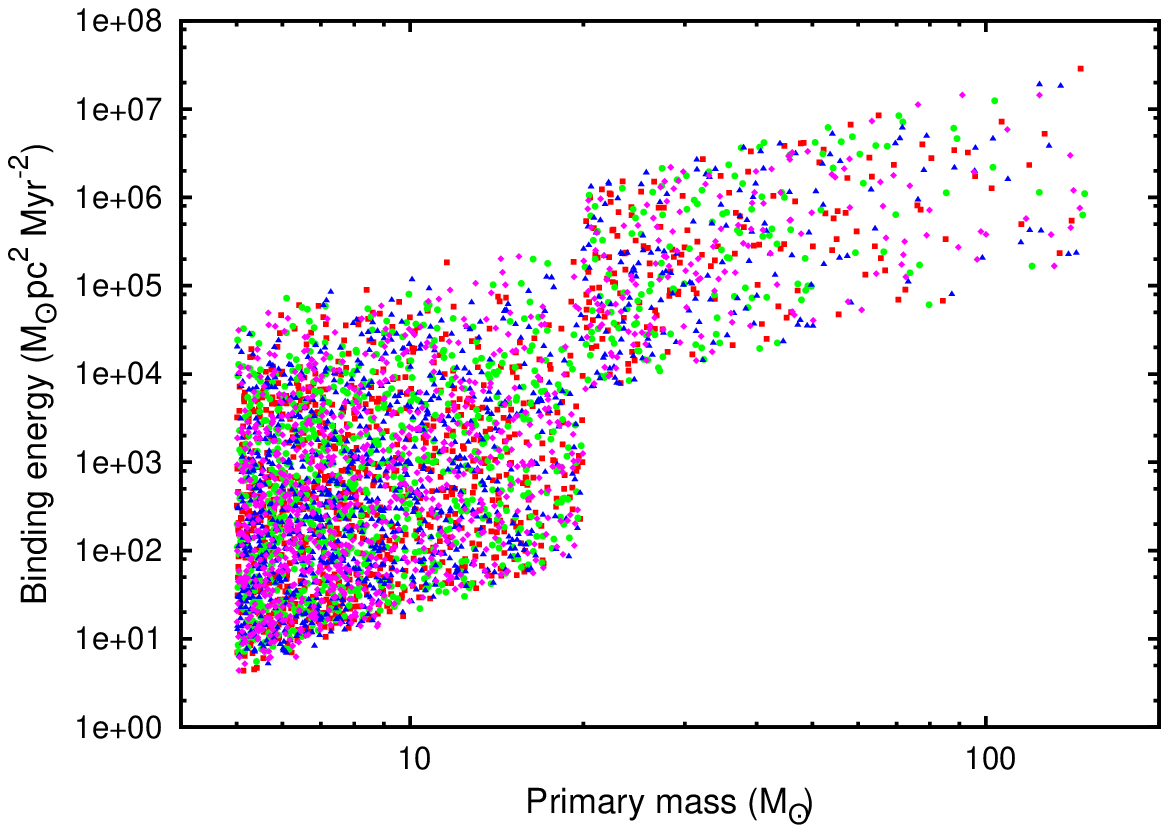}
\caption{The binding energies of the initial binaries in our computed models
as a function of their primary masses, where the two groups of binaries are clearly distinguishable
across $20\Ms$. The binaries from the 4 initial models (see Sec.~\ref{nbint}) are superimposed
which are distinguished by the different symbols.
Binaries with primary masses $>20\Ms$ are assigned significantly smaller and narrowly
distributed orbital periods than the other group (see Sec.~\ref{initcond}) making the former ones significantly
harder. The range of binding energies compares well with that of observed O-star binaries \citep{sev2010}.}
\label{fig:EvM}
\end{figure}

To understand these trends, we first note that an ejected single star of mass $M$ and velocity $V$
can appear in two main ways, \viz, (i) it is ejected due to a kinetic energy (K.E.) boost in a close, super-elastic encounter
\citep{hg75} with a hard binary where no exchange of members has occurred (a flyby encounter)
and (ii) it is ejected from a hard binary being replaced by a more massive star (an exchange encounter).
In both cases, the K.E. of ejection is of the order of the binary's binding energy, on average.
The trend of the lower boundary can be understood by noting that a single star is
likely to experience a flyby in a strong encounter
with a hard binary (most of them have nearly equal mass components in our model) only if the binary components
are more massive than the intruder, since otherwise the latter is likely to get trapped in the binary by exchanging with one
of its (lighter) members. In other words, to recoil a mass $M$ in a flyby encounter, one needs
a binary with components of mass $m_s\gtrsim M$. 
Similarly, a star of mass $M$ can preferentially be ejected from a binary only by being replaced
by an intruder of mass $m_s\gtrsim M$.
Since the ejected star, in both cases, will typically have a K.E. of the order
of the binary's binding energy \citep{hg75,hhm96}, we have $MV^2\sim m_s^2/a$, where $a$ is the binary's semi-major-axis.
Hence the condition $m_s \gtrsim M$ implies $V^2\gtrsim M/a$; the smallest possible ejection velocity increases
with $M$, for a given $a$. Hence, the widest (hard) binaries in our model give rise to the lower boundary of
the scatter in Fig.~\ref{fig:mvplot} (lower panel). Indeed, the eye-estimated lower limit of the data for the lower mass ejecta
can be well represented by a curve $V\propto M^{1/2}$ shown by the pink line segment in
Fig.~\ref{fig:mvplot} (lower panel). This is also true for the massive end; their lower boundary can be represented 
by another curve (the green line segment) but with the same proportionality. The discontinuity is expected due
to that in the binary period distribution (\cf Sec.~\ref{initcond}; Fig.~\ref{fig:EvM}). 

On the other hand, the upper boundary of the plot is formed by the
ejecta generated by binaries having the highest binding energies. Fig.~\ref{fig:EvM} shows that the hardest binaries
are also the most massive ones, as a result of our chosen primordial binary distribution (see Sec.~\ref{initcond}).
They can therefore eject stars over the entire mass range with K.E. of the order of their binding energy
forming the upper boundary that is independent of $M$. Note that as an order of magnitude estimate,
the highest binding energies are $\sim 10^6 - 10^7 \Ms {\rm~pc}^2 {\rm~Myr}^{-2}$ which yield
$V \sim 10^{2.5} - 10^3 {\rm~km} {\rm~s}^{-2}$. This indeed agrees with the highest $V$s in
Fig.~\ref{fig:mvplot} in terms of order of magnitude.

In Fig.~\ref{fig:mvplot} (lower panel) it can be seen that the median of the ejection velocities (thick black line)
tend to $V \approx 50$ km s$^{-1}$ for the most massive runaways and the corresponding quartiles (thin black lines) span between
about $40 - 60$ km s$^{-1}$. Taking into account the outliers with $V$ smaller than the first quartile
in the massive end of the $M-V$ plot, these ejection velocities are similar to the inferred true velocity
of VFTS 682 \citep{blh2011}. This result, therefore, already
strongly indicates that VFTS 682 is a typical runaway VMS from R136. The highest $V$ reached
in our computations, however, exceeds 300 km s$^{-1}$ as can be read from Fig.~\ref{fig:mvplot}.

We further inspect that all of our computations eject VMSs with kinematic properties
agreeing fairly with those of VFTS 682, \viz, $V \approx 40$ km s$^{-1}$, $R \approx 30$ pc and
$M \approx 150\Ms$. Such instances are shown in Table~\ref{tab:tab1}. 
These results dictate that the VFTS 682 is an expected, very likely runaway
VMS from R136.

As additional information to the reader and for convenience of comparison with observations,
we provide the data corresponding to Fig.~\ref{fig:mvplot} in the form of an online table (Table~\ref{tab:online2}).
In addition to the
3-dimensional velocities and the (instantaneous) masses of all the runaways from all of our computations, we also provide
the corresponding line-of-sight and tangential velocities and projected positions.
For the aid of the reader in doing further analyses, we also provide the corresponding components of the
positions and the velocities
in an additional online table (Table~\ref{tab:online3}).

\subsection{Another runaway: 30 Dor 016}\label{dor16}

Another notable VMS in the 30 Doradus is 30 Dor 016 which is also thought to be a runaway
from the central R136 cluster \citep{evns2010}. This star, located at $\approx 120$ pc projected distance from R136,
has a radial velocity of $\approx 85$ km s$^{-1}$
and its mass is inferred to be $\approx 90\Ms$ (see \citealt{evns2010} and references therein). The possibility of
\#016 having a close companion being quite unlikely as these authors conclude based on VFTS multi-epoch
spectroscopy, it is very likely a runaway from R136. Given that this VMS
is about 1 Myr old \citep{evns2010}, its projected distance implies a minimum of $\approx 120$ km s$^{-1}$ tangential velocity.
Combining this with its radial velocity, \#016 has at least $\approx 150$ km s$^{-1}$  
3-dimensional velocity. Two of our computations are indeed found to eject runaways with
similar properties which are shown in Table~\ref{tab:tab2}.

\subsection{Mass-dependence of runaway stars}\label{ejctfrac}

\begin{figure}
\centering
\vspace{-0.5cm}
\includegraphics[angle=0,height=9cm]{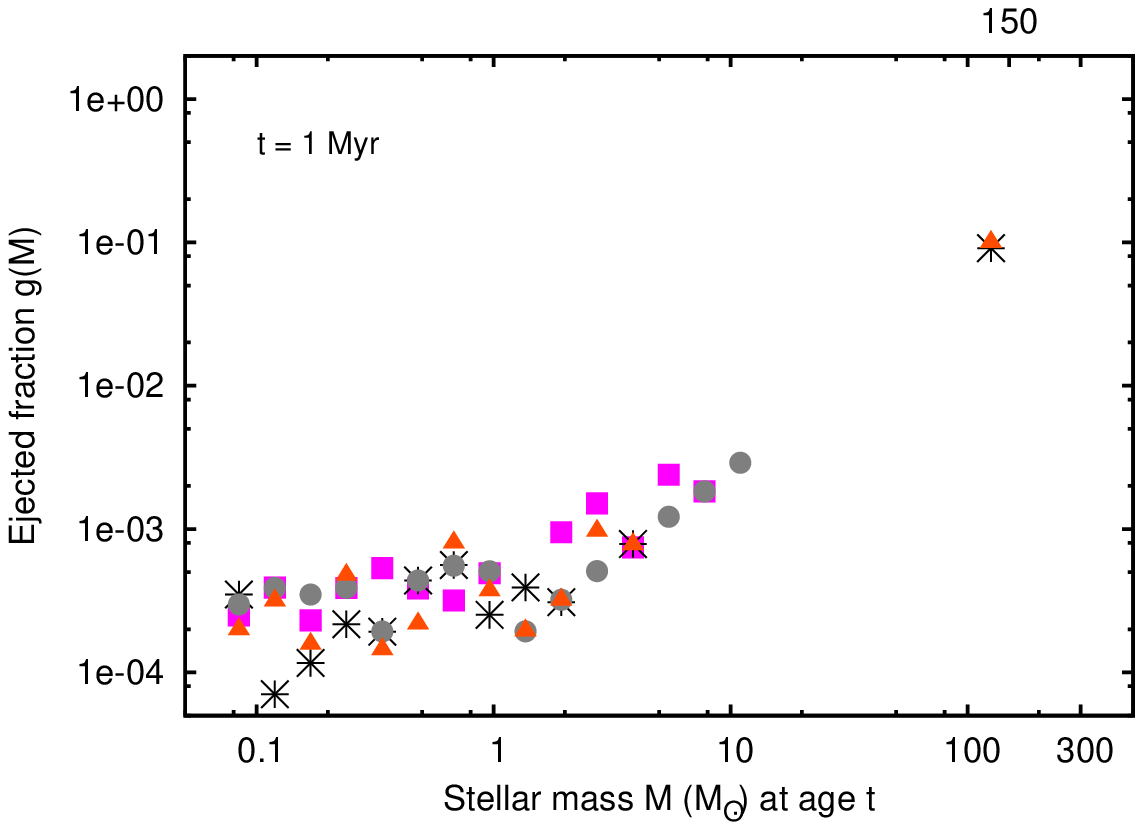}
\includegraphics[angle=0,height=9cm]{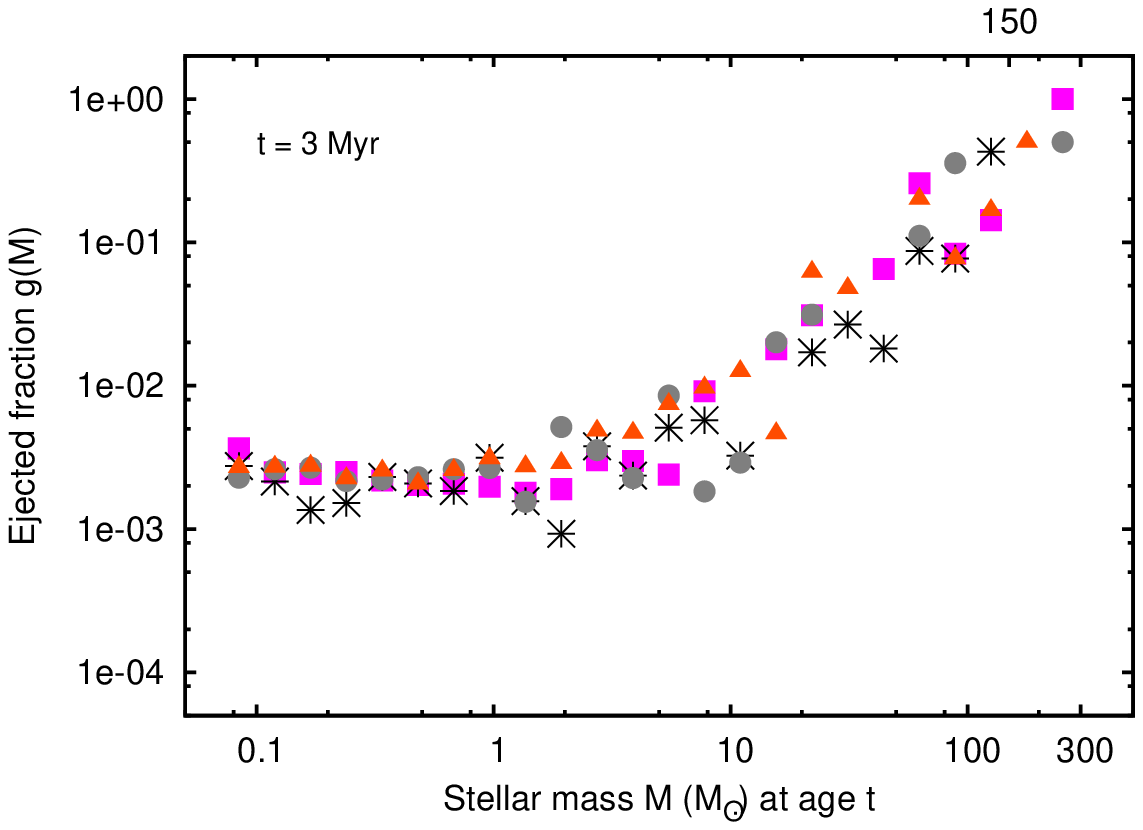}
\vspace{-0.7cm}
\caption{The runaway or ejected fraction of stars $g(M)$ as a function of stellar mass $M$ at $t=1$ Myr 
(top) and 3 Myr (bottom). The masses are divided into 25 equal bins in the logarithmic scale over the range
$0.05\Ms < M < 300\Ms$.
The outcomes from all the computations, which are distinguished by the different
symbols, are overplotted in each panel. 
For $M \gtrsim 5\Ms$, $g(M)$ increases considerably and becomes nearly unity for the most massive bin. This
indicates that the dynamical ejection process becomes increasingly important with increasing stellar mass such that
the most massive members are nearly all ejected. The flatness of $g(M)$ for $M \lesssim 5\Ms$ may be an
artifact of the primordial binary population adapted in the computations. See text for details.}
\label{fig:ejfrac}
\end{figure}

It is very interesting to study the mass (or spectral class) dependence of the runaway members
as it directly points to the pure dynamical nature of the ejection process. Although the number
of stars bound to the cluster decreases strongly with increasing stellar mass, more massive stars and binaries are
more centrally concentrated due to mass segregation and hence interact with each other more efficiently and frequently.
This causes the runaway fraction of stars (defined as the ratio of the number of stars, within a bin around mass $M$,
moving away from the cluster at $R>10$ pc to the total number of stars in the whole system, \ie, including all the
bound and the ejected members, within the same mass bin)
to increase with the stellar mass as shown in Fig.~\ref{fig:ejfrac}, where the outcomes
of all the computations are overplotted at $t=1$ Myr and 3 Myr. In Fig.~\ref{fig:ejfrac}, the ejected fraction $g(M)$
increases considerably with $M$ for $M \gtrsim 5\Ms$ but remains nearly flat and small for lower $M$. This is probably
an artifact of our initial binary population where only the stars with $m_s > 5\Ms$ are in binaries and therefore
are efficient in ejecting only those members which are more massive than $5\Ms$. Current technology does not allow
us to perform direct N-body integrations of R136-type clusters in which all stars are initially in binaries. Such
computations are available only for moderate mass clusters \citep{pk98,pketal2001}.

\section{Discussions and outlook}\label{discuss}

In the present work, we have computed the dynamical evolution of model clusters whose structure conform with the observed
global properties of R136 --- the central massive cluster in the 30 Dor complex of the LMC,
using the direct N-body integration method. The evolution of the individual
stars, chosen initially from the canonical IMF with the standard $150\Ms$ upper cutoff \citep{wk2004},
has also been incorporated and as well the evolution of the individual binaries. We focus on the
ejection of massive stars from our model clusters which is a process that depends crucially
on the properties of the primordial binaries in the cluster. For computational simplicity,
we have taken only the stars with initial masses $m_s > 5\Ms$ to be in binaries (see Sec.~\ref{initcond}) which are the only ones
that are efficient in ejecting the massive stars. Hence, the properties of the massive runaways
are not expected to be affected significantly by the absence of lower mass binaries.

Recent observations of the R136 and the 30 Dor region have raised fundamental questions
regarding massive star formation mechanisms. In particular, the apparently isolated
and relatively slow-moving single VMS VFTS 682 has raised
suspicion that it might be an instance of isolated massive star formation (\citealt{blh2011}; see Sec.~\ref{intro}).
Additionally, the presence of single VMSs in R136 with inferred initial masses upto $\approx 320\Ms$ has
questioned the canonical $150\Ms$ upper limit of the IMF \citep{crw2010}. 

\begin{figure}
\centering
\vspace{-1.0cm}
\includegraphics[angle=0,height=8cm]{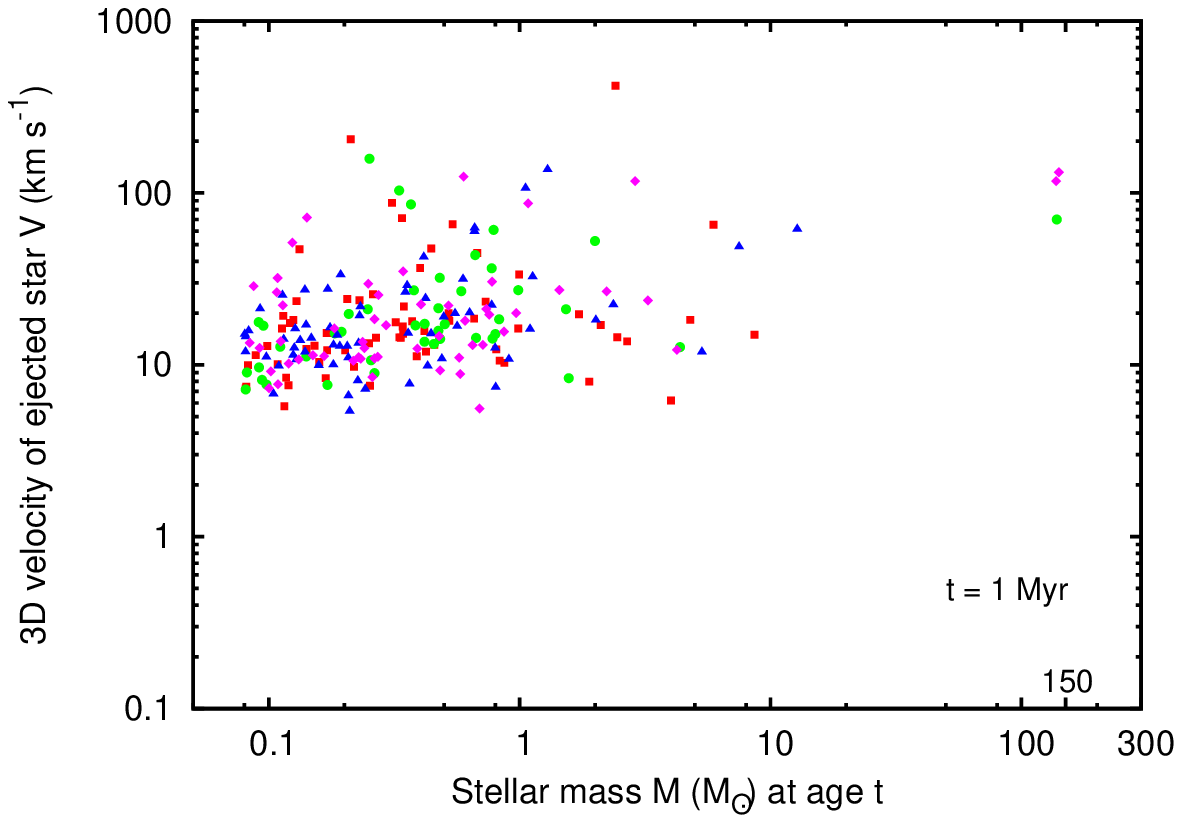}
\includegraphics[angle=0,height=8cm]{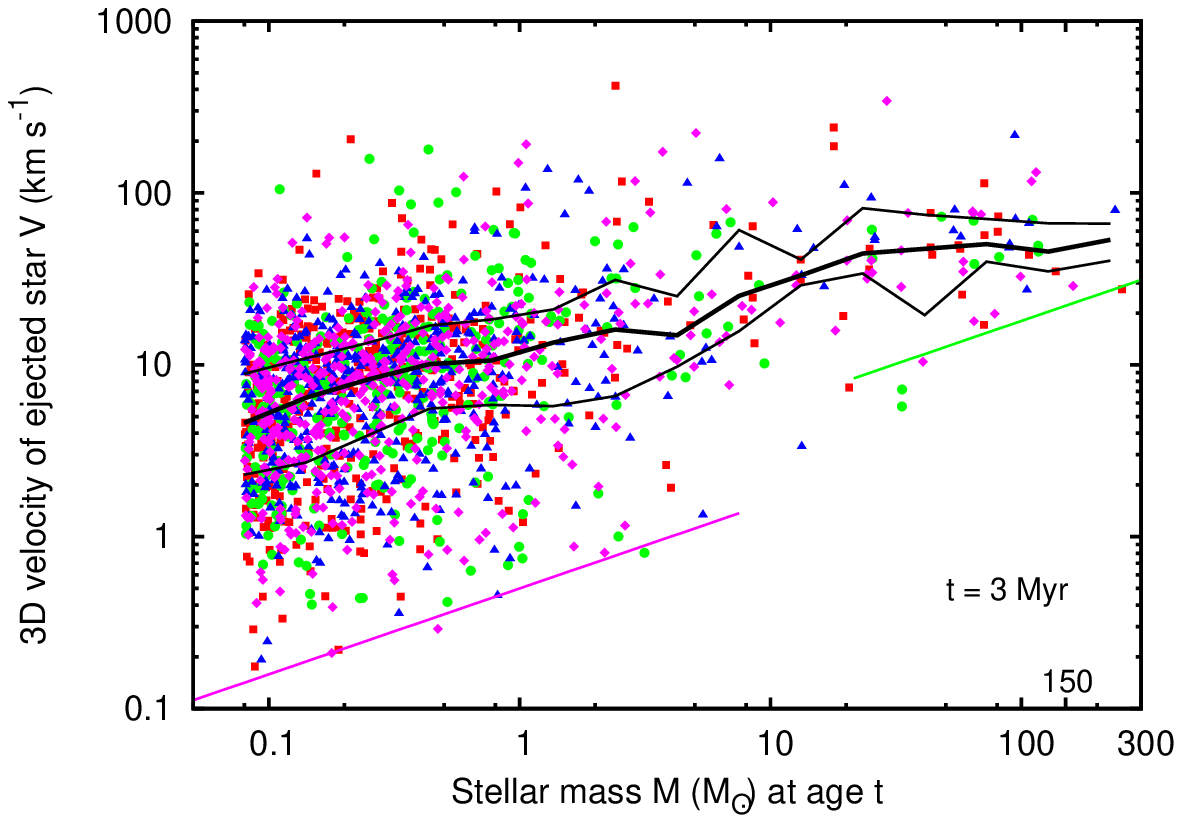}
\vspace{-0.8cm}
\caption{The 3-dimensional velocity $V$ vs. the mass $M$ of all the runaway single stars (filled symbols) at $t=1$ Myr (top)
and 3 Myr (bottom) as obtained from our calculations.
Different symbols represent the outcomes from different computations. The thick black line in the bottom panel
is the median of the scatter along the $V$ axis and the thin black lines are the corresponding first and the third quartiles.
In computing these percentiles, the data-points are divided into 15 equal bins in the logarithmic scale over the range
$0.05\Ms < M < 300\Ms$.
At the massive end, the scatter lies within the quartile range of about 40 - 60 km s$^{-1}$ (bottom panel) in striking
similarity with the inferred true velocity of VFTS 682. The lower boundary to the majority of the data-points
can be well represented by the power-law $V\propto M^{1/2}$, which are the slim pink and the green line segments (bottom panel).
The percentiles and the lower boundaries are constructed only for the bottom panel ($t=3$ Myr) as the
data in the top panel ($t=1$ Myr) is much sparser. See text for further details.}
\label{fig:mvplot}
\end{figure}

The most important
outcome of our computations is the confirmation that a ``slow runaway'', with a 3-dimensional velocity
similar to that of VFTS 682, is in fact the most probable type of ejected VMS from a R136-like
cluster (\cf Fig.~\ref{fig:mvplot}; Sec.~\ref{vfts682}). In fact, all of our computed models
yield one or more runaways with kinematic properties agreeing fairly with those of VFTS 682 (\cf Table~\ref{tab:tab1}).
Given such a likeliness of a VFTS 682-type runaway from R136, this apparently isolated star clearly does \emph{not} imply
isolated massive star formation and it is very likely a former member of R136.

Furthermore, as explained in Sec.~\ref{res}, massive, close binaries are necessary to dynamically eject VMSs from
star clusters, which, in turn, are susceptible to merge due to their hardening and/or eccentricity-pumping, by the
frequent close encounters that they receive. As our computations show (see Sec.~\ref{res}),
such massive binary mergers can easily produce single stars,
within a few Myr, with masses well exceeding the $150\Ms$ upper limit, even if the cluster begins with the canonical
upper limit. Our 4 computations have produced upto $\approx 250\Ms$ members and merger products upto $\approx 300\Ms$
are possible if the most massive binaries merge. Therefore, it doesn't seem to be a surprise that R136 has
upto $\approx 320\Ms$ single-star members, given the large uncertainties in the stellar evolution models
used to infer the masses (see \citealt{crw2010} and references therein) and the presence of these super-canonical VMSs (or the
presence of a super-saturated mass function) is \emph{not}
an indication that R136 was born with an IMF having an upper limit that substantially exceeds the widely accepted $150\Ms$
limit. These VMSs can as well be massive binary mergers.

It is, of course, important to note that there are aspects in our assumptions and initial conditions
that favor ejection of VMSs and formation of massive merger products. First, the assumption of no mass-loss
during MS-MS mergers (see Sec.~\ref{nbint}) is an idealization. Although SPH computations indicate that
no mass is lost during a MS-MS collision \citep{sil2001}, the studied collisions are typically between low mass stars
and such mass conservation is not necessarily true for massive MS-MS mergers. Furthermore, the merged MS
star can spin rapidly, even beyond break-up, in which case the net mass gain would be much smaller or nil.
Second, the assumption of a complete mixing (see Sec.~\ref{nbint}) is also an idealization; it is possible
that the denser helium-rich core(s) preferably sink to the center of the merger product resulting only
a partial mixing of unburnt hydrogen with the helium core. This would lead to a partial refuelling or rejuvenation
of the merged MS stars's core and hence a smaller lifetime. In other words, the treatment of complete mixing maximizes
the lifetime of the merged star and hence its probability of getting dynamically ejected and/or being 
observationally detected. Although idealized, these conditions are those which can currently be adapted at best
in a direct N-body calculation; other widely-used direct N-body codes such as the ``NBODY6++'' \citep{spur99}
and the ``STARLAB'' \citep{pz2001} also utilize similar synthetic stellar evolution and merger recipes. 

Third, the VMS ejection and the massive merger formation are also facilitated by our adapted
initial or primordial mass segregation (see Sec.~\ref{initcond}). While the motivation for choosing
this condition is indeed to maximize these effects, primordial mass segregation is inferred to be
true for several Galactic globular clusters \citep{bg2008,mrk2010,strd2011} and open clusters
(\eg, \citealt{bonn98,lit2003,chen2007}).
An important outlook would be to study the effects of
a varying degree of initial mass segregation.

Another ingredient that causes ejection of runaway VMSs and the presence of $m_s > 150\Ms$ VMSs in our
computations likely, is the presence of relatively close primordial binaries beyond $m_s > 20\Ms$ (see Sec.~\ref{initcond}).
However, such a binary distribution is in accordance to the observational fact that O-stars are generally found
in close binaries. Notably, the range of the O-star binaries' initial orbital periods in our models is taken
to be similar to the observed one \citep{sev2010}.
In this context, recall that we adapt a $P$-distribution that follows
a single \"Opik's law and also take the binary mass-ratios ($q$) close to unity (see Sec.~\ref{initcond}).
However, the $P$-distribution of
O-stars in O-rich clusters is observed to follow a bi-uniform distribution in $\log_{10}P$ with the break
at $P\approx 10$ days; there is a significant overabundance of binaries (comprising 50\%-60\% of the O-star binary population)
for $P\lesssim 10$ days (see Eqn.~5.2 of \citealt{sev2010}). Furthermore, their mass-ratios are found to
be uniformly distributed between $0.2\leq q \leq1.0$ \citep{sev2010}.
A very important development would be to incorporate such more realistic
distributions of O-star binaries' orbital periods and mass-ratios. In fact, the \citet{sev2010} $P$-distribution would
even more strongly favor the ejection of VMSs and the formation of super-saturated stars
than the presently adapted distribution due to the overabundance of
$P\lesssim 10$ days binaries in the former distribution.
This would, of course, be at least partially counter-balanced by the $q$-distribution;
the smallest $q$ gives a binary with a binding energy smaller by a factor of a few
for the same primary-mass and semi-major-axis. 
It is, therefore, an essential outlook to study the net outcome of such effects. However, note that
the gross conclusions, as obtained from the present models, are unlikely to get affected by these details
since the range of $P$ and hence the range of binary binding energies are indeed taken to be similar.   

Finally, note that the chosen masses of the initial Plummer clusters are the upper mass limit of that of R136
($\approx10^5\Ms$; \citealt{crw2010}). A lower mass limit of R136 is $\approx 5\times10^4\Ms$ \citep{wk2004}. This being
only by a factor of two smaller than the upper limit, we do not expect that the results presented here would be affected
significantly had we chosen the lower mass limit as the total initial cluster mass. 

A corollary of our above two results is that runaway VMSs and super-canonical VMSs
(with masses exceeding the canonical upper-limit)
would coexist, in general. Of course, as explained in Sec.~\ref{res}, the latter type of VMSs can easily be ejected from their
host clusters and may be found only as runaways. It would be worthwhile to search for other massive
(and young enough) star clusters, in addition to R136, where such super-canonical VMSs exist and the clusters
can be related to VMS runaways at the same time. Incidentally, the VMS $\zeta$ Pup ($\approx 70\Ms$),
whose kinematics indicate that it is a runaway from the Vela R2 association, has recently been inferred to
be a merger product of at least two massive stars \citep{van2011}.

The main drawback of this study is the incompleteness of the primordial binary population.
As already stressed earlier, the absence of initial binaries for $m_s < 5\Ms$ is anyway unlikely to significantly influence
the ejected stellar population for $M \gtrsim 5\Ms$ which justifies such a truncation for our purposes.
The truncated primordial binary population, of course, has effects on less massive ejecta which are
pointed out in Sec.~\ref{res}. The purpose of eliminating the lower mass binaries is, of course, computational ease.
The presence of binaries substantially diminishes the speed 
of direct N-body computations (even on GPU platforms; see Sec.~\ref{nbint})
and such computations with a full spectrum of primordial binaries (\ie, 100\% primordial binary fraction) of models
of the mass that we use in this study (\ie, $N \approx 170000$; see Sec.~\ref{nbint}) are currently prohibitive
even with the fastest available GPUs and workstation processors. Nevertheless, we wish to emphasize
again that our computations in this study are the largest sized direct N-body computations
of initially fully mass-segregated clusters in which all massive stars are in binaries reported so far,
thanks to the remarkable boost of computing speed provided by the recent GPUs (see \citealt{gforce}) and to the efficient
numerical algorithms of NBODY6 \citep{ar2003}. These calculations represent the first-time attempt to directly
evolve a real-sized massive star cluster.

An immediate improvement over our work is, of course, to explore a full primordial binary spectrum
and the effect of their varied distribution functions which, we expect, will become plausible
in the near future given the continuing remarkable algorithmic improvements of NBODY6 and the availability of
increasingly faster GPUs. A more accurate structural modelling of R136 is also pending once
its dimensions are more clearly settled (see Sec.~\ref{initcond}).

In conclusion, our calculations imply that the observed VMSs outside R136 and also its bound VMS members are fully
consistent with the standard paradigm of massive star formation exclusively in dense environments and following
the canonical IMF. Hence, the observations discussed here cannot be considered as instances violating these
regimes.

\acknowledgments

We are thankful to Sverre Aarseth of the Institute of Astronomy, Cambridge, U.K., for his effort in
the continuing remarkable improvements of NBODY6 and his gracious help during our computations.
We are also thankful to Keigo Nitadori of RIKEN, Japan, for his developments that resulted in the
exemplary hardware-acceleration of NBODY6. We thank the anonymous referee whose suggestions have
improved the quality of the manuscript substantially.

\begin{deluxetable}{ccccc}
\tablenum{1}
\tablecaption{A list of runaway single stars with properties fairly close to those of VFTS 682, as
obtained from our computations. The columns are as follows: Col.~(1): model number, Col.~(2): evolutionary
time $t$ at which the runaway is detected, Col.~(3): mass $M$ of the runaway at $t$, Col.~(4): its distance
$R$ from the cluster's center of density and Col.~(5): its 3-dimensional velocity $V$. $M$, $R$ and $V$ of these
runaway stars agree
fairly with those estimated observationally for VFTS 682 (last line; \citealt{blh2011}).}
\label{tab:tab1}
\tablehead{
\colhead{model number} & \colhead{time $t$ (Myr)}
& \colhead{mass $M$ ($\Ms$)} & \colhead{distance $R$ (pc)} & \colhead{velocity $V$ (km s$^{-1}$)}
}
\startdata
    1        &   2.8   &    256.4   &  31.9   &  27.5\\
             &   3.2   &    135.9   &  26.6   &  34.8\\
\hline
    2        &   2.6   &    126.4   &  27.7   &  45.7\\   
             &   2.6   &    125.9   &  29.9   &  49.4\\   
\hline
    3        &   2.6   &    106.9   &  45.7   &  27.3\\
\hline
    4        &   1.9   &    169.1   &  29.3   &  29.0\\
             &   1.9   &    116.9   &  35.2   &  32.8\\
\hline
VFTS 682     & $<3.0$  &    $\approx 150.0$   &  $\approx30.0$   &  $\approx 40.0$
\enddata
\end{deluxetable}

\begin{deluxetable}{cccccccccc}
\tabletypesize{\small}
\tablenum{2}
\tablecaption{Data for all the runaway stars as obtained from all of our computations at
1 Myr and 3 Myr. The columns are as follows: Col.~(1): mass $M$ of the runaway at 1 Myr/3 Myr,
Col.~(2): its tangential velocity $v_t$ (all velocities are relative to the cluster's center of mass)\tablenotemark{a},
Col.~(3): line-of-sight or radial velocity $v_{\rm rad}$\tablenotemark{b}, Col.~(4): 3-dimensional velocity $V$,
Col.~(5): its projected distance $r$ from the cluster's center of density, Col.~(6): $1\Rightarrow$ single,
$2\Rightarrow$ binary and Col.~(7): identity of the star $I$.
}
\label{tab:online2}
\tablehead{
\colhead{$M$ ($\Ms$)} & \colhead{$v_t$ (km s$^{-1}$)} & \colhead{$v_{\rm rad}$ (km s$^{-1}$)} & \colhead{$V$ (km s$^{-1}$)}
& \colhead{$r$ (pc)} & \colhead{single/binary} & \colhead{$I$}
}
\startdata
   5.9 &  60.0 & -25.9 &  65.3 	& 15.3 &   1 &     1626\\
   8.6 &  10.4 &  10.7 &  15.0 	&  9.1 &   1 &     960\\
   0.2 &   8.4 &   8.8 &  12.2 	&  7.9 &   1 &     4491\\
   0.4 &  15.8 &  44.8 &  47.5 	&  5.8 &   1 &     6332\\
   0.3 &   6.8 &   3.3 &   7.6 	&  9.6 &   1 &     8042\\
  40.5 &   6.5 &  -8.2 &  10.4 	& 19.5 &   2 &     170968\\
  94.0 &  50.3 & 209.9 & 215.8 	& 83.7 &   1 &     10\\
 236.3 &  31.3 & -72.8 &  79.2 	&  9.4 &   1 &     5\\
\enddata
\tablenotetext{a}{The reference frame is chosen fixed and originated at the cluster's center of mass.
Its axes are taken to be oriented arbitrarily as there is no preferred directionality due to the absence of an external
field. All the projections are taken on the $x-y$ plane.}
\tablenotetext{b}{The velocity component normal to the plane of projection or the $z$ velocity component is simply
taken to be the radial velocity due to the arbitrariness in the choice of the coordinate axes. The corrections due
to the spatial extent is negligible for the distance of R136.}
\tablecomments{The full Table~\ref{tab:online2} can be accessed online in the electronic edition
of the journal (see http://tinyurl.com/3kdy64g for a text version).
Examples of its data are shown here to indicate its format.}
\end{deluxetable}

\begin{deluxetable}{ccccccccc}
\tabletypesize{\scriptsize}
\tablenum{3}
\tablecaption{Data for all the runaway stars as obtained from all of our computations at
1 Myr and 3 Myr. The columns are as follows: Col.~(1): mass $M$ of the runaway at 1 Myr/3 Myr,
Col.~(2): its identity $I$, Col.~(3-5): components of its position relative to the cluster's
center of density\tablenotemark{c}, Col.~(6-8): components of its velocity relative to the cluster's
center of mass, Col.~(9) $1\Rightarrow$ single, $2\Rightarrow$ binary.}
\label{tab:online3}
\tablehead{
\colhead{$M$ ($\Ms$)} & \colhead{$I$} & \colhead{$x$ (pc)} & \colhead{$y$ (pc)} & \colhead{$z$ (pc)} &
\colhead{$v_x$ (km s$^{-1}$)} & \colhead{$v_y$ (km s$^{-1}$)} & \colhead{$v_z$ (km s$^{-1}$)} & \colhead{single/binary}
}
\startdata
   5.9 & 1626 	&          -12.6 &     -8.7 	 &  -6.5 	&  -49.7 	&  -33.6 	&  -25.9 &  1 \\
   8.6 & 960 	&           -6.9 &      5.9 	 &   9.0 	&   -8.0 	&    6.7 	&   10.7 &  1 \\
   0.2 & 4491 	&            4.4 &      6.6 	 &   8.8 	&    3.4 	&    7.6 	&    8.8 &  1 \\
   0.4 & 6332 	&           -2.2 &      5.3 	 &  16.4 	&   -6.5 	&   14.5 	&   44.8 &  1 \\
   0.3 & 8042 	&           -6.8 &     -6.7 	 &   4.5 	&   -4.7 	&   -4.9 	&    3.3 &  1 \\
  40.5 & 170968 &          -16.1 &    -11.1 	 & -25.2 	&   -5.4 	&   -3.5 	&   -8.2 &  2 \\
 107.2 & 170726 &           10.7 &    -55.2 	 &  47.6 	&    6.3 	&  -32.7 	&   28.2 &  2 \\
 236.3 & 5 	&           -8.8 &     -3.4 	 & -94.9 	&   -7.8 	&  -30.3 	&  -72.8 &  1 \\
\enddata
\tablenotetext{c}{The reference frame is chosen fixed and originated at the cluster's center of mass.
Its axes are taken to be oriented arbitrarily as there is no preferred directionality due to the absence of an external
field.}
\tablecomments{The full Table~\ref{tab:online3} can be accessed online in the electronic edition
of the journal (see the URL http://tinyurl.com/6fepz6b for a text version).
Examples of its data are shown here to indicate its format.}
\end{deluxetable}

\begin{deluxetable}{ccccc}
\tablenum{4}
\tablecaption{A list of runaway single stars with $M$, $R$ and $V$ similar to those of 30 Dor 016
(last line; \citealt{evns2010} and references therein), as obtained from
our computations. The meanings of the columns are the same as in Table~\ref{tab:tab1}.}
\label{tab:tab2}
\tablehead{
\colhead{model number} & \colhead{time $t$ (Myr)}
& \colhead{mass $M$ ($\Ms$)} & \colhead{distance $R$ (pc)} & \colhead{velocity $V$ (km s$^{-1}$)}
}
\startdata
    3        &    3.1    &    90.9   &   103.3    &   144.9   \\
\hline
    4        &    1.3    &    138.6  &   122.0    &   131.8   \\
\hline
30 Dor 016   & $\approx 1.0$ & $\approx 90$ &  $\approx 120$ & $\approx 150$     
\enddata
\end{deluxetable}

\end{document}